\documentclass[12pt]{article}
\usepackage{setspace} 
\usepackage[dvips]{graphicx}
\usepackage{graphics}
\usepackage{amssymb}
\usepackage{bm}
\usepackage{amsmath}
\usepackage{color}
\usepackage{epstopdf}
\usepackage[margin=1in]{geometry}

\usepackage[utf8]{inputenc}
\begin{document}

\begin{center}
{\bf Essay written for the Gravity Research Foundation 2026 \\
Awards for Essays on Gravitation}
\end{center}

\begin{center}
{\bf A No-Go Theorem for Topological Bridges with Matter-Vacuum Coupling
}
\end{center}

\begin{center}
{Rodrigo Maier\footnote{rodrigo.maier@uerj.br}}
\end{center}

\begin{center}
Departamento de F\'isica Te\'orica, Instituto de F\'isica, \\Universidade do Estado do Rio de Janeiro,\\
Rua S\~ao Francisco Xavier 524, Maracan\~a,\\
CEP20550-900, Rio de Janeiro, Brasil
\end{center}

\begin{center}
{(February 25, 2026)}
\end{center}

\begin{abstract}
Traversable topological bridges traditionally require exotic matter, violating the Null Energy Condition (NEC). This essay investigates whether matter-vacuum coupling can circumvent this necessity. Focusing on zero-tidal-force solutions, we establish a rigorous no-go theorem for static configurations, proving that such coupling cannot bypass the requirement for NEC violation. We demonstrate that the geometric flare-out condition is incompatible with NEC-compliant sources, regardless of the coupling $Q$ or equation of state. Crucially, the vacuum fails to shield the throat; instead, interaction gradients mathematically obstruct the required geometry. This result suggests that causality protection is inherent in the field equations, rendering the vacuum’s evolution a regulator rather than a facilitator of topological shortcuts, thereby reinforcing the robustness of classical energy conditions.
\end{abstract}

\newpage

Since the seminal work of Morris and Thorne\cite{Morris:1988cz}, the theoretical possibility of traversable wormholes has remained one of the most provocative features of General Relativity. These topological bridges represent a fundamental departure from the trivial connectivity of flat spacetime, yet they appear to be protected by a rigorous censorship mechanism. Specifically, the Einstein field equations demand that the throat of a traversable bridge must satisfy a geometric flare-out condition, which later was understood\cite{visser} as a definitive requirement for the violation of the Null Energy Condition (NEC). This necessity for exotic matter represents a significant ontological barrier. Such matter typically falls outside the scope of the Standard Model and threatens the stability of the semi-classical vacuum by allowing for the existence of closed timelike curves\cite{Morris:1988tu}.

Recent research in general wormhole physics has undergone a significant paradigm shift, moving from abstract existence proofs toward rigorous astrophysical phenomenology and stability analysis within modified gravity frameworks. A central pillar of this evolution is the Generalized Ellis-Bronnikov (GEB) model\cite{Kar:1995jz}, a sophisticated extension of the seminal 1973 solution\cite{Bronnikov:1973fh}. While the original metric was characterized by a symmetric, static geometry, modern GEB models utilize a more versatile shape function $b(r)$ and redshift function $\Phi(r)$. These generalizations allow physicists to describe compact objects that lack an event horizon -- thereby avoiding the information paradox associated with black holes -- while maintaining finite tidal forces suitable for theoretical traversability. Recent investigations have demonstrated that the inclusion of angular momentum in these generalized metrics can freeze the radial instabilities that plagued earlier versions, suggesting that rotating Bronnikov-type structures could realistically persist in a dynamic universe\cite{Azad:2024axu}. Beyond internal stability, researchers have turn their attention to identify observational fingerprints of these structures, particularly through the gravitational wave (GW) memory effect. This effect refers to the permanent displacement of spacetime following the passage of a gravitational radiation pulse. Studies in GEB spacetimes indicate that the absence of a singularity and the presence of a throat create a backscattering of waves that differs fundamentally from the signatures produced by black holes\cite{Molla:2024loc}. This distinct memory signature, alongside gravitational lensing echoes, provides a tangible method for future interferometers to differentiate a GEB wormhole from a standard Kerr black hole. In the original 1973 Bronnikov model, exotic matter (specifically a phantom scalar field with negative kinetic energy) was strictly required to provide the outward pressure needed to keep the throat open. However, in these recent generalized models, there is a concerted effort to minimize or eliminate this requirement. By using modified gravity theories, such as $f(R)$ or Einstein-Gauss-Bonnet gravity, the exotic effects are often transferred to the geometry of spacetime itself rather than a physical substance, potentially allowing these wormholes to exist without violating standard energy conditions\cite{Ernazarov:2025fyv,Godani:2023jhq}. Nevertheless,
while these models allow the physical matter fluid to potentially remain NEC-compliant, they often introduce new degrees of freedom, such as the Ostrogradsky ghost\cite{Woodard:2006nt}, or dynamical instabilities\cite{DeFelice:2010aj,Cuyubamba:2018jdl} that complicate the physical interpretation of the resulting spacetime.

Parallel to these geometric modifications, a sophisticated alternative has emerged from observational cosmology. Recent results from the Dark Energy Spectroscopic Instrument (DESI)\cite{DESI:2024uvr,Pan:2025qwy} have uncovered tensions in the standard $\Lambda$CDM model, with data release analyses favoring a dynamical or interacting dark energy over a static cosmological constant at significances reaching up to $4.2\sigma$. A particularly compelling framework for this is the Running Vacuum Model (RVM), extensively developed by Solà and collaborators\cite{Moreno-Pulido:2020anb,SolaPeracaula:2021gxi}. Unlike the static $\Lambda$ term, the RVM treats the vacuum energy density $\rho_{vac}$ as a dynamic quantity fundamentally linked to the evolution of the universe\cite{Gomez-Valent:2024tdb}. By introducing a coupling function $Q$ that governs the energy-momentum exchange between matter and the vacuum, these models offer potential resolutions to the coincidence problem and modern cosmological tensions such as the $H_0$ and $\sigma_8$ discrepancies\cite{SolaPeracaula:2021gxi,Sola:2017znb,Wang:2016lxa}.

The motivation for considering this as a potential resolution for topological bridges is profound. If a non-trivial energy-momentum exchange exists between the vacuum and matter sectors on a cosmological scale, the vacuum component may function locally as a topological buffer, mediating the stress-energy requirements of the bridge. The tantalizing prospect is that the vacuum’s inherently negative pressure might be leveraged via the coupling $Q$ to bear the burden of the flare-out condition, effectively shielding the matter from the topological demands of the geometry. If the vacuum component can absorb the NEC-violating requirements of the flare-out condition, the surrounding matter fluid doesn't have to be exotic at all. This leads to a fundamental question: If the vacuum energy density is dynamic and coupled to matter, as suggested by recent cosmological data, can the negative energy needed for a wormhole be sourced from the vacuum's own evolution, leaving the baryonic matter perfectly normal?

In this essay, we rigorously examine this potential circumvention. We establish a definitive no-go theorem within the physically favorable class of zero-tidal-force solutions for the case of static and spherically symmetric configurations. We demonstrate that the NEC violation required by the flare-out condition persists regardless of the chosen coupling function or the fluid’s equation of state. The persistence of the NEC violation confirms that vacuum interactions cannot offset the geometric demands of the throat, solidifying the vacuum's role as a safeguard against the formation of shortcuts in General Relativity.

To start we consider the gravitational field equations in the presence of an interacting vacuum component, 
\begin{eqnarray}
\label{eqefe}
G^{\mu\nu} = \kappa^2 \left( T^{\mu\nu} + T_{\text{vac}}^{\mu\nu} \right),    
\end{eqnarray}
where $G^{\mu\nu}$ is the Einstein tensor and the total stress-energy is partitioned into a matter fluid $T^{\mu\nu}$ and a vacuum contribution $T_{\text{vac}}^{\mu\nu}$. From Bianchi identities then follows that
\begin{eqnarray}
\nabla_\mu (T^{\mu\nu} + T_{\text{vac}}^{\mu\nu}) = 0.    
\end{eqnarray}
In the framework of interacting vacuum component\cite{vanderWesthuizen:2025rip} we introduce a non-minimal coupling via an energy-momentum transfer 4-vector $Q^\nu$. This allows for a dynamical exchange between the sectors:
\begin{eqnarray}
\label{eqc1}
\nabla_\mu T^{\mu\nu} &=& -Q^\nu, \\
\label{eqc2}
\nabla_\mu T_{\text{vac}}^{\mu\nu} &=& Q^\nu.
\end{eqnarray}
As mentioned above this phenomenological approach is widely utilized to address cosmological tensions and explore the stability of dark sector interactions\cite{Epelbaum:2025mxo,Wang:2024vmw}.

In order to examine the problem of topological bridges in the case of static and spherically symmetric configurations
let us consider a geometry described by the line element
\begin{eqnarray}
\label{geo}
ds^2=-e^{2\Phi(r)}dt^2+\frac{1}{1-b(r)/r}dr^2+r^2(d\theta^2+\sin^2\theta d\phi^2),    
\end{eqnarray}
where $\Phi(r)$ is the redshift function and $b(r)$ is the shape function\cite{Morris:1988cz}. It can be easily shown that the direct substitution of the metric coefficients into the Einstein tensor yields a purely diagonal structure. By virtue of the field equations, the effective energy-momentum tensor $T^{\mu\nu} + T_{\text{vac}}^{\mu\nu}$ 
must inherit this diagonal form. In a general framework we shall then assume that the matter content is given
by an anisotropic fluid whose energy-momentum tensor reads
\begin{eqnarray}
\label{tmunu}
T^{\mu}_{~~\nu}={\rm diag}(-\rho, -\tau, p, p),    
\end{eqnarray}
where $\rho(r)$ is the energy density while $\tau(r) = -p_r(r)$ is the radial tension (the negative of the radial pressure). The term $p(r)$ corresponds to the tangential pressure. Regarding the vacuum counterpart, we generalize the case of a constant cosmological term\cite{Lemos:2003jb} by incorporating a spatially dependent function $V(r)$, defined as follows:
\begin{eqnarray}
\label{tmunuvac}
T^{\mu\nu}_{\text{vac}}=-V(r)g^{\mu\nu}.    
\end{eqnarray}
From Einstein field equations (\ref{eqefe}) we then obtain:
\begin{eqnarray}
\label{eqtt}
&&\frac{b^\prime}{r^2}=\kappa^2(\rho+V),\\
\label{eqrr}
&&\frac{b}{r^3}-2\Big(1-\frac{b}{r}\Big)\frac{\Phi^\prime}{r}=\kappa^2(\tau+V),\\
\label{eqthetatheta}
&&\Big(1-\frac{b}{r}\Big)\Big[\Phi^{\prime\prime}+{\Phi^\prime}^2-\Big(\frac{b^\prime r-b}{2r^2(1-b/r)}\Big)\Big(\Phi^\prime+\frac{1}{r}\Big)+\frac{\Phi^\prime}{r}\Big]=\kappa^2 (p-V).
\end{eqnarray}
where primes denote derivatives with respect to $r$.

As we are restricting ourselves to the case of static configurations, from now on we shall assume that the energy-momentum transfer $4$-vector $Q^\mu$ is defined through a general coupling function $Q(r)$. That is:
\begin{eqnarray}
Q^\mu= Q(r) \delta^\mu_{~r}.    
\end{eqnarray}
In the following we shall denote $Q$ by the transfer function.
Therefore, from (\ref{eqc1})--(\ref{eqc2}) it follows
\begin{eqnarray}
\label{tp}
&&\tau^\prime=(\rho-\tau)\Phi^\prime-\frac{2}{r}(p+\tau)+ Q,\\    
\label{vp}
&&V^\prime=-Q.    
\end{eqnarray}

In this essay we shall consider the physically favorable class in which static interior observers measure
zero tidal forces so that $\Phi={\rm constant}$ (and $\Phi^\prime\equiv 0$). In this case, equations (\ref{eqtt}) and (\ref{eqrr}) give us
\begin{eqnarray}
\label{ri}
\rho(r)=\tau+\frac{r b^\prime-b}{\kappa^2 r^3}.    
\end{eqnarray}
Equations (\ref{eqtt}) and (\ref{eqthetatheta}) on the other hand furnish
\begin{eqnarray}
\label{pi}
p(r)=-\tau-\Big(\frac{rb^\prime-3b}{2\kappa^2r^3}\Big).    
\end{eqnarray}
The substitution of (\ref{ri}) and (\ref{pi}) in (\ref{eqtt})--(\ref{eqthetatheta}) yields
\begin{eqnarray}
\label{ti}
\tau(r)=\frac{b}{\kappa^2 r^2}-V.    
\end{eqnarray}
For $\Phi={\rm constant}$ it is then easy to see that (\ref{ri})--(\ref{ti}) automatically satisfy (\ref{eqtt})--(\ref{eqthetatheta}) together with (\ref{tp})--(\ref{vp}), leaving the shape and transfer functions undetermined. 
However, at this stage we draw the reader's attention to a word to note.
In the standard framework of Lorentzian wormholes with vanishing tidal forces the shape function is usually given 
ad hoc and the requirement for exotic matter independently arises as a direct consequence of the flare-out condition at the throat, which imposes the violation of the NEC. By considering that the gravitational source is not a single isolated fluid, but rather a composite system wherein baryonic matter interacts non-minimally with a vacuum component one might assume that such interaction could provide an effective energy-momentum tensor ${\cal T}^{\mu\nu}=T^{\mu\nu}+T_{\text{vac}}^{\mu\nu}$
whose effective energy density and radial tension do not violate the NEC.
By defining ${\cal T}^{\mu\nu}=T^{\mu\nu}+T_{\text{vac}}^{\mu\nu}$ it is then easy to see
through a simple inspection of (\ref{tmunu}), (\ref{tmunuvac}) and (\ref{ri}) that
\begin{eqnarray}
\label{nectacl}
-({\cal T}^{t}_{~t}-{\cal T}^{r}_{~r})\equiv\tau-\rho\equiv\frac{b-rb^\prime}{\kappa^2 r^3},
\end{eqnarray}
As we will see in the following, this result constitutes a direct violation of the NEC once one imposes the flare-out condition. 

It is well known that the flare-out condition -- which constrains the shape function -- comes from pure geometrical grounds. In the present context we may obtain it by embedding the geometry (\ref{geo}) -- with $t = \text{const.}$ and $\theta = \pi/2$ -- into a three-dimensional Euclidean spacetime. The line element of this two-dimensional surface is given by:
\begin{eqnarray}
\label{tb}
ds^2 = \frac{1}{1 - b/r} dr^2 + r^2 d\phi^2.    
\end{eqnarray}
To visualize this slice, we represent it as a surface $z = z(r)$ within a 3D Euclidean space. In cylindrical coordinates $(z, r, \phi)$, the Euclidean metric is expressed as $ds^2 = dz^2 + dr^2 + r^2 d\phi^2$. By applying the chain rule to the vertical displacement, the metric on the surface becomes
\begin{eqnarray}
\label{eu}
ds^2 = \left[ \left( \frac{dz}{dr} \right)^2 + 1 \right] dr^2 + r^2 d\phi^2.    
\end{eqnarray}
Comparing (\ref{tb}) and (\ref{eu}) we can determine the specific slope of the embedding surface:
\begin{eqnarray}
\frac{dz}{dr} = \pm \left( \frac{r}{b} - 1 \right)^{-1/2}.    
\end{eqnarray}
To be a solution of a topological bridge the geometry (\ref{tb}) must have a minimum radius $r_0$ that acts as the throat of the wormhole. In the immediate vicinity of this throat, the geometry must satisfy the flare-out condition
which ensures that the the second derivative of the inverse embedding function $r(z)$ to be positive, namely,
\begin{eqnarray}
\label{fo}
\frac{d^2r}{dz^2} = \frac{b - rb^\prime}{2b^2} > 0.    
\end{eqnarray}
However, according to (\ref{nectacl}) the flare-out condition can be rewritten as
\begin{eqnarray}
\label{foc}
\frac{d^2r}{dz^2} = \frac{\kappa^2r^3}{2b^2} (\tau-\rho) > 0,    
\end{eqnarray}
which implies $-({\cal T}^{t}_{~t}-{\cal T}^{r}_{~r})>0$ or simply $\tau>\rho$, thus violating the NEC. 
Physically, this result exposes a crucial mechanical failure in the coupling hypothesis. As $V'(r) = -Q(r)$, any attempt to use the coupling to ``inject'' negative energy into the throat is counteracted by the spatial gradient of the vacuum itself. The interaction effectively ``chokes'' the throat's geometry, forcing the total stress-energy complex back into a NEC-violating requirement, regardless of the transfer function's form.

The persistent inability of interacting vacuum components to alleviate violations of the NEC suggests that the Topological Censorship Conjecture is significantly more robust than previous explorations of static manifolds indicated. As established by Friedman et al.\cite{Friedman:1993ty}, the structural prohibition against shortcuts in the spacetime fabric is not merely a localized artifact of simplified General Relativity, but rather a persistent feature that remains invariant even when transitioning to semi-classical or dynamic vacuum modifications. By demonstrating that the introduction of energy exchange between vacuum and matter fields fails to soften the geometric requirements of the throat, we uncover a profound resilience in the causal structure of the universe, reinforcing the notion that non-trivial topology remains fundamentally inaccessible under standard energy constraints\cite{Galloway:1996hx}.

If zero-tidal-force solutions are strictly prohibited by the NEC within this framework, we are led to accept this definitive no-go theorem for macroscopic interstellar transit. While our theorem is mathematically rooted in the $\Phi' = 0$ class, physical constraints suggest this is the only case relevant to viable transit. Any theoretically permissible wormhole geometry divorced from the benign zero-tidal-force assumption would necessitate steep redshift gradients to satisfy the flare-out condition. This shift in the metric profile introduces catastrophic physical consequences: any hypothetical traveler would be subjected to an intense gravitational blueshift, wherein ambient background radiation is compressed into a lethal ionizing spectrum, rendering the throat biologically impenetrable. Simultaneously, the departure from the vanishing tidal force condition mandates the presence of extreme gravitational shears that would likely exceed the mechanical integrity of any known baryonic structure. Thus, the zero-tidal-force case is not merely a simplification, but the only interesting domain for sourcing traversable shortcuts.

Furthermore, these findings illuminate a fundamental tension between the geometric requirements of traversable wormholes and the foundational motivations within cosmology and quantum field theory. In contemporary cosmology, an interacting vacuum is frequently introduced to address the cosmic coincidence problem or to mitigate persistent tensions in cosmological datasets, typically modeled via the coupled conservation laws (\ref{eqc1})--(\ref{eqc2}). However, the inherent rigidity required for such a vacuum to drive late-time cosmic acceleration appears to be the same property that precludes the localization of the negative energy densities necessary for throat stabilization. Crucially, this result implies that the very cosmological solutions currently invoked to resolve dark energy tensions -- such as the RVM -- cannot be ``repurposed'' to create stable local topology. The physics that drives the expansion of the universe appears to be fundamentally distinct from the physics required to puncture its connectivity.

In the domain of quantum field theory, the necessity for an interacting vacuum to maintain a macroscopic throat encounters the rigorous barrier of Quantum Energy Inequalities (QEIs). As established by Ford and Roman\cite{Ford:1995wg}, these inequalities dictate that the sampled energy density along a worldline is bounded below:
\begin{eqnarray}
\int_{-\infty}^{\infty} \langle T_{\mu \nu} k^{\mu} k^{\nu} \rangle f(t) dt \geq -\frac{C}{t_0^4}    
\end{eqnarray}
where $f(t)$ represents a sampling function with a characteristic timescale $t_0$. These constraints imply that any localized manifestation of negative energy density must be followed by a significantly larger, compensating positive energy flux. This phenomenon, often termed ``quantum kickback'', acts to effectively choke the wormhole throat, preventing the maintenance of a stable bridge. This behavior aligns with the Average Null Energy Condition (ANEC), as discussed by Fewster and Roman\cite{Fewster:2005gp}, which requires that the integrated energy density along a null geodesic remains non-negative:
\begin{eqnarray}
\int \rho_{eff} d\lambda \geq 0.    
\end{eqnarray}
Consequently, the very quantum and cosmological mechanisms invoked to explain the large-scale structure of the universe act as a prohibitive regulatory framework against the localized topological deformations required for traversability.

The persistence of NEC violations in the presence of an interacting vacuum component marks a significant boundary in theoretical physics. This result highlights a sobering hierarchy within gravitational laws: the geometry of the wormhole throat is an uncompromising master. The failure of $Q$-coupling to fix the NEC violation implies that the protection of chronology and causality is ``baked into'' the Einstein tensor itself, regardless of how we slice the energy-momentum side of the equation. Regardless of the complexity of the vacuum’s internal interactions, the stress-energy complex remains a servant to the fundamental flare-out laws of spacetime. The geometry does not merely invite the presence of exotic matter; it demands it as an absolute toll, reinforcing the idea that gravity acts as its own most rigorous censor, prioritizing causal consistency and the integrity of the vacuum over topological flexibility.


\begin{thebibliography}{99}

\bibitem{Morris:1988cz}
M.~S.~Morris and K.~S.~Thorne,
Am. J. Phys. \textbf{56}, 395-412 (1988).

\bibitem{visser} M. Visser, {\it Lorentzian Wormholes: From Einstein to Hawking} AIP Press (1995).

\bibitem{Morris:1988tu}
M.~S.~Morris, K.~S.~Thorne and U.~Yurtsever,
Phys. Rev. Lett. \textbf{61}, 1446-1449 (1988).

\bibitem{Kar:1995jz}
S.~Kar, S.~Minwalla, D.~Mishra and D.~Sahdev,
Phys. Rev. D \textbf{51}, 1632-1638 (1995).

\bibitem{Bronnikov:1973fh}
K.~A.~Bronnikov,
Acta Phys. Polon. B \textbf{4}, 251-266 (1973).


\bibitem{Azad:2024axu}
B.~Azad, J.~L.~Bl{\'a}zquez-Salcedo, F.~S.~Khoo and J.~Kunz,
Phys. Rev. D \textbf{109}, no.12, 124051 (2024)
[arXiv:2403.08387 [gr-qc]].

\bibitem{Molla:2024loc}
N.~U.~Molla, H.~Chaudhary, U.~Debnath, G.~Mustafa and S.~K.~Maurya,
Eur. Phys. J. C \textbf{85}, no.1, 15 (2025)
[arXiv:2406.09492 [gr-qc]].

\bibitem{Ernazarov:2025fyv}
K.~K.~Ernazarov,
Int. J. Theor. Phys. \textbf{64}, no.10, 256 (2025)
[arXiv:2504.06223 [gr-qc]].

\bibitem{Godani:2023jhq}
N.~Godani and S.~Kala,
Int. J. Mod. Phys. D \textbf{32}, no.10, 2350067 (2023).



















\bibitem{Woodard:2006nt}
R.~P.~Woodard,
Lect. Notes Phys. \textbf{720}, 403-433 (2007)
[arXiv:astro-ph/0601672 [astro-ph]].

\bibitem{DeFelice:2010aj}
A.~De Felice and S.~Tsujikawa,
Living Rev. Rel. \textbf{13}, 3 (2010)
[arXiv:1002.4928 [gr-qc]].

\bibitem{Cuyubamba:2018jdl}
M.~A.~Cuyubamba, R.~A.~Konoplya and A.~Zhidenko,
Phys. Rev. D \textbf{98}, no.4, 044040 (2018)
[arXiv:1804.11170 [gr-qc]].

\bibitem{DESI:2024uvr}
A.~G.~Adame \textit{et al.} [DESI],
JCAP \textbf{04}, 012 (2025)
[arXiv:2404.03000 [astro-ph.CO]].

\bibitem{Pan:2025qwy}
S.~Pan, S.~Paul, E.~N.~Saridakis and W.~Yang,
Phys. Rev. D \textbf{113}, no.2, 023515 (2026)
[arXiv:2504.00994 [astro-ph.CO]].

\bibitem{Moreno-Pulido:2020anb}
C.~Moreno-Pulido and J.~Sola,
Eur. Phys. J. C \textbf{80}, no.8, 692 (2020)
doi:10.1140/epjc/s10052-020-8238-6
[arXiv:2005.03164 [gr-qc]].

\bibitem{SolaPeracaula:2021gxi}
J.~Sol{\`a} Peracaula, A.~G{\'o}mez-Valent, J.~de Cruz Perez and C.~Moreno-Pulido,
EPL \textbf{134}, no.1, 19001 (2021)
[arXiv:2102.12758 [astro-ph.CO]].


\bibitem{Gomez-Valent:2024tdb}
A.~Gomez-Valent and J.~Sol{\`a} Peracaula,
Astrophys. J. \textbf{975}, no.1, 64 (2024)
[arXiv:2404.18845 [astro-ph.CO]].

\bibitem{Sola:2017znb}
J.~Sol{\`a}, A.~G{\'o}mez-Valent and J.~de Cruz P{\'e}rez,
Phys. Lett. B \textbf{774}, 317-324 (2017)
doi:10.1016/j.physletb.2017.09.073
[arXiv:1705.06723 [astro-ph.CO]].


\bibitem{Wang:2016lxa}
B.~Wang, E.~Abdalla, F.~Atrio-Barandela and D.~Pavon,
Rept. Prog. Phys. \textbf{79}, no.9, 096901 (2016)
[arXiv:1603.08299 [astro-ph.CO]].

\bibitem{vanderWesthuizen:2025rip}
M.~van der Westhuizen, A.~Abebe and E.~Di Valentino,
Phys. Dark Univ. \textbf{50}, 102121 (2025)
[arXiv:2509.04496 [gr-qc]].

\bibitem{Epelbaum:2025mxo}
E.~Epelbaum, J.~Gegelia and U.~G.~Mei{\ss}ner,
Phys. Rev. D \textbf{112}, no.10, 106005 (2025)
[arXiv:2506.19182 [hep-th]].

\bibitem{Wang:2024vmw}
B.~Wang, E.~Abdalla, F.~Atrio-Barandela and D.~Pav{\'o}n,
Rept. Prog. Phys. \textbf{87}, no.3, 036901 (2024)
[arXiv:2402.00819 [astro-ph.CO]].

\bibitem{Lemos:2003jb}
J.~P.~S.~Lemos, F.~S.~N.~Lobo and S.~Quinet de Oliveira,
Phys. Rev. D \textbf{68}, 064004 (2003)
[arXiv:gr-qc/0302049 [gr-qc]].

\bibitem{Friedman:1993ty}
J.~L.~Friedman, K.~Schleich and D.~M.~Witt,
Phys. Rev. Lett. \textbf{71}, 1486-1489 (1993)
[erratum: Phys. Rev. Lett. \textbf{75}, 1872 (1995)]
[arXiv:gr-qc/9305017 [gr-qc]].

\bibitem{Galloway:1996hx}
G.~Galloway and E.~Woolgar,
Class. Quant. Grav. \textbf{14}, L1-L7 (1997)
[arXiv:gr-qc/9609007 [gr-qc]].

\bibitem{Ford:1995wg}
L.~H.~Ford and T.~A.~Roman,
Phys. Rev. D \textbf{53}, 5496-5507 (1996)
[arXiv:gr-qc/9510071 [gr-qc]].

\bibitem{Fewster:2005gp}
C.~J.~Fewster and T.~A.~Roman,
Phys. Rev. D \textbf{72}, 044023 (2005)
[arXiv:gr-qc/0507013 [gr-qc]].





\end{thebibliography}
\end{document}